\documentclass[12pt]{article}
\usepackage[english]{babel}
\usepackage{graphicx,graphics}
\font\bss=cmr12 scaled\magstep 0
\newcommand {\vT} {v _ {\scriptscriptstyle T}}
\newcommand {\di} {\displaystyle}

\newcommand {\p} {\partial}

\begin{document}
\title{Piecewise continuous distribution function method:
  Fluid equations and  wave disturbances at stratified gas}
\author{\bss Vereshchagin D.A., S. B. Leble\\
\small Theoretical Physics Department,
\small Kaliningrad State University, \\
\small 236041, Kaliningrad, Al. Nevsky str. 14. \\
\small   Theoretical Physics and Mathematical Methods Department,\\
\small  Technical University of Gdansk, ul, Narutowicza 11/12,
Gdansk,  Poland,\\
\small  leble@mifgate.pg.gda.pl \\  \\[2ex] }
\maketitle

\renewcommand{\abstractname}{\small Abstract}
 \begin{abstract}
Wave disturbances of a stratified gas are studied. The description
is built on a basis of the Bhatnagar -- Gross -- Krook (BGK)
kinetic equation which is reduced down the level of fluid
mechanics. The double momenta set is introduced inside a scheme of
iterations of the equations operators, dividing the velocity space
 along and opposite gravity field direction.
At both half-spaces the local equilibrium is supposed. As the
result, the momenta system is derived. It reproduce Navier-Stokes
and Barnett equations at the first and second order in high
collision frequencies. The homogeneous background limit gives
the known results obtained by direct kinetics applications by
Loyalka and Cheng as the recent higher momentum fluid mechanics
results of Chen,  Rao and Spiegel. The ground state declines from
exponential at the Knudsen regime. The WKB solutions for
  ultrasound in exponentially stratified medium are constructed in explicit form, evaluated and
  plotted.
\end{abstract}

\section{ Introduction.}

                        \qquad     \qquad     \qquad  \qquad           {\bf Devoted to
                                                    D.A. Vereshchagin
                                                    memory}
                                                    \medskip

 There are
 gas dynamics problems at which it is necessary to use a
 basis going out of traditional
Navier - Stokes  hydrodynamics. It is connected to a break of the condition:
$Kn = {\it l}/L < < 1 $,
where $Kn $ -  Knudsen number, $ {\it l} $ -   free particle path, and
$L $ - characteristic scale of non-homogeneity of a problem. Perhaps, the
first work, in which a wave disturbance in a gas was investigated
from the point of view of more general kinetic approach was
the work of Wang Chang and Uhlenbeck \cite{chang}.  The
  authors have offered a method of a dispersion
relation construction  in a homogeneous gas directly from
Boltzmann equation.

  The further
theoretical and experimental researches
\cite{meyer}~-~\cite{banan} on sound propagation in a homogeneous
gas  have shown, that at Knudsen numbers of the unit order the
waves behavior considerably  differs from ones predicted on a
basis of  Navier - Stokes equations. These researches revealed two
essential features: first, the perturbations keep wave properties
at more large values of $Kn $, than it could be assumed on a basis
of the classical hydrodynamic description. Secondly, at $Kn \ge 1
$ such concepts as a wave vector and frequency of a wave become
ill-determined. May be the most adequate results that reproduce
experiments \cite{meyer} almost in all the range were obtained in
\cite{loyalka}. It is  more difficult to explore the case, when
the Knudsen number  is non-uniform in space or in time and passes
the Knudsen regime area. The statement and the solution of such
problems should definitely be based on a kinetic equations or
their advance model analogues.

Quite recently interest to the problems has grown again in connection with general
 fluid mechanics development
 \cite{leble2, veresc2, veresc3, CRS, CRS1}.
It was pushed by more deep understanding of perturbation theory
(so-called nonsingular perturbations), see, e.g. \cite{LWG}.

In his paper we consider the gas medium, stratified exponentially in gravity field, directed along z axis.
In means that the Knudsen number also depends on z: $Kn(z)$. We continue to develop the method \cite{veresc3} that
 goes up to the pioneering paper of Lees \cite{lees}.
The construction of  analytical solutions  of the model  kinetic equation
Bhatnagar -- Gross -- Krook (BGK) is extracted via separate representation of the distribution
function as the local equilibrium one but with different momenta sets at positive and negative velocity component
 $v_z$ subspaces.

Thus, the set of parameters determining a state of the gas
increases twice. Such number of parameters of the distribution
function (\ref {q6}) results in that the distribution deviates
from a local-equilibrium and accordingly widen hydrodynamics. In
the range of small Knudsen numbers $ {\it l}<< L $ we have $ \hat
M_n^+ = \hat {M}_n^-$ and distribution function~(\ref{q6}) passes
to local equilibrium one, giving a solution of the Navier-Stokes
hydrodynamical regime. For big Knudsen numbers the formula~(\ref
{q6}) gives a solution of so-called collisionless problems.
Similar ideas have resulted  successfully in a series of problems.
For example, in papers \cite{lees}~-~\cite{schidl} a method of
discontinuous distribution functions  was used for the description
of a flat and cylindrical (neutral and plasma) flows
~\cite{lees}~-~\cite{schidl}. For a flat problem the surface of
break in space of speeds was determined by the same natural
condition $V_z=0 $, and in a cylindrical case $V_r=0 $, where $V_z
$ and $V_r $,  vertical and radial component of speed of particles
respectively. The problem  of a disturbance launched by a pulse
movement of plane~\cite{schidl} was solved similarly.
 In a problem  of a shock
wave structures~\cite{schidl,{mott}, nambu} the solution  was
represented as a combination of two locally equilibrium functions,
one of which determines the function before front of a wave, and
another - the tail. In a problem of condensation and evaporation
of drops of any size~\cite{sampson, ivchenko} a  break surface was
determined by so-called  "cone of influence", thus all particles
were divided  to two types: flying "from a drop" and flying "not
from a drop".

At the first two sections we derive the basic equations using the iterations in the evolution operator
along the idea of the nonsingular perturbation method. Next (Sec. 4) we analyze the transition to a
 limiting case of  a gas disturbances at large collision frequencies up to the Barnett case.  Next (Sec. 5)
 we check the free molecular flow limit, demonstrating the declinations from the exponential behavior of the
 "atmosphere" gas density \cite{veresc4, ...}. At the final section we construct
  solutions of the main momenta system by the method VKB, considering the wave scale less than the scale of
  the inhomogeneity.

\section{ Linearized BGK equation }

 The kinetic equation with  the model collision integral
 in  BGK  form looks like:
\begin {equation}
\label {q1} \frac {\partial f} {\partial t} + \vec v\frac {\partial f} {\partial
\vec r}- g\frac {\partial f} {\partial v_z} = \nu\left (f _ {\it l}-f\right) \,
\end {equation}
here $f $ -- distribution function of a gas, $t $ -- time, $ \vec
v $ -- velocity of a particle of the gas, $ \vec r $ -- its
coordinate vector,
$$
f_{\it l}=\frac{n}{\pi^{3/2}\vT^3}     
\exp\left(-\frac{(\vec v-\vec U)^2}{\vT^2}\right)
$$
-- local-equilibrium distribution function, $H $ -- a scale of
inhomogeneity (in atmospheric  models $H =kT/mg$), $ \vT
=\sqrt{2kT/m} $ -- average thermal velocity of movement of
particles of gas, $ \nu =\nu_0\exp (-z/H) $ -- effective frequency
of collisions between particles of gas at height $z $. It is
supposed, that density of gas is denoted as  $n $, its average
speed $ \vec U = (u_x, u_y, u_z) $ and temperature $T $ are
functions of time and coordinates.

Considering small gas  perturbations, we shall approximate a
distribution function $f $ as:
$$
f(t,\vec r,\vec v)=f_0(z,\vec v)(1+\varphi(t,\vec r,\vec v))\ , \hspace{20mm}
|\varphi|<<1\ .
$$
Here $f_0 $ -- equilibrium Maxwell -- Boltzmann distribution
function , and $ \varphi $ -- dimensionless small-amplitude
perturbation to equilibrium distribution function. For
thermodynamic  parameters of a gas  we shall write:
$$
n=n_0(1+n')\ , \hspace{1cm} \vec U=\vec U'\vT\ , \hspace{1cm} T=T_0(1+T')\ ,
$$
where $n'$, $\vec U'$ and $T' $ -- dimensionless small additives
to equilibrium density, average speed and temperature gas
respectively. Expanding the local equilibrium function $f_{\it l}
$ up to the first order by small amplitudes and taking into
account, that equilibrium function $f_0 $ identically satisfies
the kinetic equation ~(\ref{q1}) we obtain linearized equation
BGK.

We shall consider one-dimensional case of the equation:
\begin{equation}
\label{q2} \frac{\partial \varphi }{\partial t}+v_z\frac{\partial \varphi }{\partial
z}-g\frac{\partial \varphi }{\partial v_z}=\nu \left\{\sum_{n=1}^3
M_n(t,z)\chi_n(\vec v)-\varphi \right\}\ .
\end{equation}

Here $ \chi_n(\vec v) $ -- eigen functions of the linearized
collisions operator:
\begin{equation}
\label{q3}
\begin{array}{rclrrcl}
\chi_1&=&1\ ,& \chi_4&=& \di \frac{2}{\sqrt5}
\frac{v_z}{\vT}\left(\frac{v^2}{\vT^2}-
\frac52\right)\ ,\vspace{2mm} \\
\chi_2&=&\di \sqrt2\frac{v_z}{\vT}\ ,&
\chi_5&=&\di \frac1{\sqrt3\vT^2}\left(v^2-3v_z^2\right)\ ,\vspace{2mm}\\
\chi_3&=&\di \sqrt{\frac23}\left(\frac{v^2}{\vT^2}-\frac32\right)\ ,& \chi_6&=&\di
\sqrt{\frac 65}\frac {v_z}{\vT^3}\left(v^2-
\frac 53v_z^2\right)\ .\\
\end{array}
\end{equation}
$M_n $ -- the moments of distribution function determined through
scalar products:
\begin{equation}
\label{q4} M_n(t,z) = <\chi_n,\varphi> \equiv \frac{1}{\pi^{1/2}\vT^3}\int d\vec
v\: \exp \left(-v^2/\vT^2\right)\cdot \chi_n(\vec v)\varphi(t,z,\vec v)\ .
\end{equation}
In linear approach the moments $M_n $ are linked to the
thermodynamic variables as:
\begin{equation}
\label{q5} M_1=\frac{n-n_0}{n_0},\qquad M_2=\sqrt{2}\frac{u_z}{\vT},\qquad
M_3=\sqrt{\frac 23}\frac{T-T_0}{T_0}\ .
\end{equation}

\section{ A method of piecewise continuous distribution function
 in linear approximation.}

Following the idea of a method of piecewise continuous
distribution functions \cite{veresc4} let's search  the solution $
\varphi $ of the equations~(\ref{q2}) as a combination of two
locally equilibrium distribution functions, each of which gives
the contribution to the corresponding area of the velocity
subspace as follows:
\begin{equation}
\label{q6} \varphi=\left\{
\begin{array}{rcl}
\varphi^+&=&\hat{M}_1^+ +\sqrt{2}\di \frac{v_z}{\vT}\cdot \hat{M}_2^+ +
\sqrt{\frac23}\left(\frac{v^2}{\vT^2}-\frac32\right)\hat{M}_3^+\ ,
\qquad v_z\ge 0\\
\varphi^-&=&\hat{M}_1^- +\sqrt{2}\di \frac{v_z}{\vT}\cdot \hat{M}_2^- +
\sqrt{\frac23}\left(\frac{v^2}{\vT^2}-\frac32\right)\hat{M}_3^-\ ,
\qquad v_z< 0\\
\end{array}
\right.
\end{equation}
The parameters $\hat M_n^{\pm}$ of locally equilibrium
distributions functions
 are linked to
 the correspondent density, average speed and temperature and three higher moments.
This idea of the method of two-fold distribution functions of
~(\ref {q6}) further is realized as follows. Multiplying the BGK
equation ~(\ref {q2}) by the eigen functions~(\ref {q3})  with the
account of~(\ref {q4}), calculating integrals we  obtain a system
of the differential equations for the moments $M_n $:
\begin{equation}
\label{q7}
\begin{array}{lll}
\smallskip
& \di \frac{\p M_1}{\p t}+\frac{\vT}{2}\frac{\p M_2}
{\p z}-\frac{\vT}{2H}M_2=0\ ,\\
\smallskip
& \di \frac{\p M_2}{\p t}+ \vT\frac{\p }{\p z}\left(M_1+M_3-\frac23 M_5\right)-
\frac{\vT}{H}\left(M_3-\frac23 M_5\right)=0\ ,\\
\smallskip
& \di \frac{\p M_3}{\p t}+\frac{\vT}{3}\frac{\p }
{\p z}\left(M_2+M_4\right)-\frac{\vT}{3H}M_4=0\ ,\\
\smallskip
& \di \frac{\p M_4}{\p t}+\vT\frac{\p }{\p z} \left(\frac52 M_3-\frac16
M_5\right)-\frac{\vT}{2H}M_5=-\nu(z) M_4
\ ,\\
\smallskip
& \di \frac{\p M_5}{\p t}+\vT\frac{\p }{\p z} \left(-M_2-\frac25 M_5+\frac95
M_6\right)+\frac{\vT}{H}\left(\frac25 M_4-
\frac95 M_6\right) =-\nu(z) M_5 
\ ,\\
\smallskip
& \di \frac{\p M_6}{\p t}+\frac{\vT}{3}\frac{\p M_5}
{\p z}+\frac{\vT}{6H}M_5=-\nu(z) M_6\ .\\
\end{array}
\end{equation}
The moments $M_n $ of the distribution function  are connected to
parameters
 $ \hat{M}_n ^\pm $ of the two-fold distribution functions~(\ref {q6}) by the relations:
\begin{equation}
\label{q8}
\begin{array}{rclrcl}
 M_1&=&\di M_1^{+}+M_2^{-}\ ,
&M_4&=&\di -M_1^{-}+\frac72 M_3^{-}\ ,\\
 M_2&=&\di \sqrt 2\left(M_1^{-}+\frac12 M_2^{+}+\frac12 M_3^{-}\right)\ ,
&M_5&=&\di -\frac1{\sqrt 3}M_2^{-}\ ,\\
 M_3&=&\di \sqrt\frac32\left(\frac{1}{3}M_2^{-}+M_3^{+}\right)\ ,
&M_6&=&\di \sqrt{\frac 65}\left(\frac13 M_1^{-}+\frac12 M_3^{-}\right)\ .\\
\end{array}
\end{equation}
Here the following notations are introduced:
$$
\begin{array}{rclrcl}
 M_1^{+}&=&\di \frac12(\hat{M}_1^{+}+\hat{M}_1^{-})\ ,
&M_1^{-}&=&\di \frac1{2\sqrt \pi}(\hat{M}_1^{+}-\hat{M}_1^{-})\ ,\\
 M_2^{+}&=&\di \frac1{\sqrt 2} (\hat{M}_2^{+}+\hat{M}_2^{-})\ ,
&M_2^{-}&=&\di \frac{1}{\sqrt{2\pi}}(\hat{M}_2^{+}-\hat{M}_2^{-})\ ,\\
 M_3^{+}&=&\di \frac1{\sqrt6}(\hat{M}_3^{+}+\hat{M}_3^{-})\ ,
&M_3^{-}&=&\di \frac1{\sqrt{6\pi}}(\hat{M}_3^{+}-\hat{M}_3^{-})\ .\\
\end{array}
$$
The feedback between parameters $M_n^{\pm} $ and the moments $M_n $ is determined
from~(\ref {q8}) and looks like:
$$
\begin{array}{rclrrl}
 M_1^{+}&=&\di M_1+\sqrt 3 M_5\ ,
&M_1^{-}&=&\di \frac{3}{10}\sqrt 5\left(-M_4+\frac76\sqrt 6 M_6\right)\ ,\\
 M_2^{+}&=&\di \sqrt 2 M_2+\frac 25\sqrt 5 M_4+
\frac {12}{5}\sqrt{\frac{10}{3}}M_6\ ,
&M_2^{-}&=&\di -\sqrt 3 M_5\ ,\\
 M_3^{+}&=&\di \left(\sqrt{\frac23} M_4+\frac{\sqrt3}{3}M_5\right)\ ,
&M_3^{-}&=&\di \frac15\sqrt 5 \left(M_4+\frac12 \sqrt6 M_6\right)\ .\\
\end{array}
$$

\section{ A limiting case of  large collision frequencies.}

Within  limits of small Knudsen numbers, from the equations~(\ref
{q7}) it follows, that $M_4, M_5, M_6<<M_1, M_2, M_3 $. Then in a
limit $\nu\to\infty$ ($Kn \ll 1$, hydrodynamical limit) we have
$M_4, M_5, M_6 \to 0 $ and the system~(\ref {q7}) tends to the
linearized  Euler's system:
$$
\begin{array}{lll}
\smallskip
& \di  M_{1t}+\frac{\vT}{2}M_{2z}-\frac{\vT}{2H}M_2=0\ ,\\
\smallskip
& \di M_{2t}+\vT\left(M_1+M_3\right)_z-\frac{\vT}{H} M_3=0\ ,\\
\smallskip
& \di M_{3t}+\frac{\vT}{3}M_{2z}=0\ ,
\end{array}
$$
where the bottom indices $t,z$ denote partial derivatives by the
time and the coordinate z.

In the next order of the perturbation theory in the small parameter
$(\nu\tau_0)^{-1}<<1$ (here $\tau_0=H/\vT$ -- characteristic time) we find
connections:
$$
M_4 =-\frac{5\vT}{2\nu} M_{3z}\ ,\hspace{1cm} M_5 = \frac{\vT}{\nu} M_{2z}\ ,
\hspace{1cm}
M_6 = 0\ ,
$$
substituting which into the system of first three equations~(\ref {q7}) one arrives at the
linearized system  of  Navier --- Stokes equations:
$$
\begin{array}{lll}
\smallskip
& \di M_{1t}+\frac{\vT}{2} M_{2z}-\frac{\vT}{2H}M_2=0\ ,\\
\smallskip
& \di M_{2t}+\vT\left(M_1+M_3\right)_z-\frac{\vT}{H} M_3
-\frac{2}{3}\frac{\vT^2}{\nu}M_{2zz}=0\ ,\\
\smallskip
& \di M_{3t}+\frac{\vT}{3} M_{2z}-\frac{5}{6}\frac{\vT^2}{\nu}M_{3zz}=0\ .
\end{array}
$$

In the higher orders of the theory from the system~(\ref {q7}) s
the linearized Barnett's equations follow. For example, in the
third order of the small parameter $(\nu\tau_0)^{-1}<<1$ we have:
$$
\begin{array}{l}
\smallskip
\di M_{1t}+\frac{\vT}{2} M_{2z}-\frac{\vT}{2H}M_2=0\ ,\\
\smallskip
 \di M_{2t}+\vT\left(M_1+M_3\right)_z-\frac{\vT}{H} M_3
-\frac{2}{3}\frac{\vT^2}{\nu}M_{2zz}-\frac{\vT^3}{\nu^2}
\left(M_{1z}+\frac{1}{H}M_{1}-\frac{1}{H}M_3\right)_{zz}=0\ ,\\
\smallskip
 \di M_{3t}+\frac{\vT}{3} M_{2z}-\frac{5}{6}\frac{\vT^2}{\nu}M_{3zz}-
\frac{\vT^3}{\nu^2}\left(M_{2z}-\frac{1}{H}M_2\right)_{zz}=0\ .
\end{array}
$$
\vskip 2mm

\section{ A free molecular  flow limit.}

Next we would study the opposite limiting case - so-called Knudsen
regime ($\nu <<1/\tau_0$). For simplicity we restrict ourselves
now by a consideration of stationary solutions of the
systems~(\ref {q7}).

 At $\nu = 0$ ( collisionless gas) the general solution of the system (\ref{q7})
 is expressed in elementary functions$$
\begin{array}{rclrcl}
 M_1&=&\di C_1+C_3\frac{z}{H}-3 C_5\exp\left(-\frac{z}{2H}\right)\ ,
&M_2&=&\di C_2\exp\left(\frac{z}{H}\right)\ ,\vspace{2mm} \\
 M_3&=&\di C_3-\frac13 C_5\exp\left(-{z}{2H}\right)\ ,
&M_4&=&\di \left(C_4-C_2\frac{z}{H}\right)\exp\left(\frac{z}{H}\right)\ ,\vspace{2mm} \\
 M_5&=&\di C_5\exp\left(-\frac{z}{2H}\right)\ ,
&M_6&=&\di \frac19(2C_4-7C_2-5C_6)\exp\left(\frac{z}{H}\right)\ .\\
\end{array}
$$
Generally, when $\nu\ne 0$, the solutions for $M_2$ and $M_4$
remain the same, while for $M_5$ it is possible to derive the
linear non-uniform equation with variable coefficients
$$
M_{5zz}+M_{5z}\frac{1}{2H}-M_5\frac{5}{3}\frac{\nu_0^2}{\vT^2}
\exp\left(-\frac{2z}{H}\right) = -\frac{7\nu_0}{3\vT H}C_2 .
$$
Leaving borders of the system motionless, let's assume that
average speed of the gas $M_2=0$. Then, if $C_2=0$,  we go to the
linear homogeneous equation for the moment $M_5$. In dimensionless
variables $z \to z/H$ it is:
\begin{equation}
\label{q9}
M_{5zz}+\frac12 M_{5z}-K e^{-2z}M_5=0\ ,
\end{equation}
where
$$
K=\frac{5}{3}\left(\frac{\nu_0H}{\vT}\right)^{2}=\frac{5}{3}(Kn)^{-2}\ .
$$

Let's change the coordinate variable  as $\exp(-2z)=t$, then the equation~(\ref {q9})
 becomes:
\begin{equation}
\label{qe10}
t M_{5tt}+\di\frac{3}{4}M_{5t}-k M_5=0\ ,
\end{equation}
where $k=\di\frac{5}{12}\frac{{{\nu}_0}^2 H^2}{{\vT}^2}.$

The equation~(\ref {qe10}) represents the linear differential
equation ( degenerate hypergeometric one) of the second order with
variable  coefficients.  The solution may be found as the
generalized power series
$$
M_5(t)= \sum\limits_{n=0}^\infty a_n^{(5)} t^n+ t^{\frac{1}{4}}
    \sum\limits_{n=0}^\infty b_n^{(5)} t^n\ .
$$
Returning to the former dimensionless variable $\bar
z\quad=\di\frac{z}{H}$ one arrives at:
\begin{equation}
\label{qe11} M_5(\bar z)=\sum_{n=0}^\infty a_n^{(5)} \exp(-2 n \bar z)+
        \exp(-\di\frac{\bar z}{2})
        \sum_{n=0}^\infty b_n^{(5)} \exp(-2 n \bar z) \ .
\end{equation}
For the coefficients $a_n^{(5)}$ and $b_n^{(5)}$ we have recurrent formulas:
$$
a_{n+1}^{(5)}=\di\frac{4k}{(n+1)(4n+3)}a_n^{(5)}\ , \qquad
b_{n+1}^{(5)}=\di\frac{4k}{(n+1)(4n+5)}b_n^{(5)}\ .
$$
Substituting $M_5$ into the equations for other moments yields:
\begin{equation}
\label{qe12}
\begin{array}{rcl}
M_2(\bar z)&=&0\ ,\vspace{2mm} \\
M_4(\bar z)&=&M_{40}\exp(\bar z)\ ,\vspace{2mm} \\
M_3(\bar z)&=&M_{30}-\di\frac{2}{5}\frac{{\nu}_0 H}{\vT} \bar z M_{40}+
        \di\frac{1}{5}a_0^{(5)} \bar z +\frac{1}{15}a_0^{(5)}+\sum_{n=1}^\infty a_n^{(3)}
        \exp\left(-2 n \bar z\right) -\vspace{2mm}\\
       &-&\di\frac{1}{3} b_0^{(5)} \exp\left(-\frac{\bar z}{2}\right)
        + \exp\left(-\di\frac{\bar z}{2}\right)
        \di\sum\limits_{n=1}^\infty b_n^{(3)}  \exp\left(-2 n \bar z\right)\ ,
        \\
\end{array}
\end{equation}
where
$$
 a_n^{(3)}=\di\frac1{30}\frac{2n-3}{n} a_n^{(5)}\ ,\qquad
 b_n^{3}=\di\frac{1}{15}\frac{4 n - 5}{4 n+1} b_n^{(5)}\ .
$$
Next the expressions for $M_6$ and $M_1$:
\begin{equation}
\label{qe13}
\begin{array}{rcl}
 M_6(\bar z)&=&M_{60} \exp(\bar z ) +\di\frac{5}{9}\frac{{\nu}_0 H}{\vT}
         \exp(\bar z)\sum_{n=0}^\infty b_n^{(6)}
         \exp\left(-\bar z(4 n+5)/2\right)
        + \vspace{2mm}  \\
         &+&\di\frac{5}{9}\frac{{\nu}_0 H}{\vT}
         \exp(\bar z)\sum\limits_{n=0}^\infty a_n^{(6)}
         \exp\left(-2\bar z(n+1)\right)     \ ,  \vspace{3mm}\\
 M_1(\bar z)&=&M_{10} +\di\frac{1}{5}\frac{{\nu}_0 H}{\vT}
         M_{40} (2 \bar z-{\bar z}^2)+\left(\di\frac{3}{5}+
         \frac{1}{10}{\bar z}^2-\di\frac{4}{5} \bar z\right) a_0
         + \vspace{2mm}  \\
         &+&\di 3 b_0 \exp\left(-\di\frac{\bar z}{2}\right)
         +M_{30}(\bar z-1)+
         \sum\limits_{n=1}^\infty a_n^{(1)} \exp\left(-2 n \bar z\right)
         + \vspace{2mm} \\
         &+&\di \exp\left(-\di\frac{\bar z}{2}\right)
         \sum\limits_{n=1}^\infty
         b_n^{(1)}\exp\left(-2 n \bar z\right)\ .\\
\end{array}
\end{equation}
The coefficients are:
$$
 a_n^{(1)}=\di\frac{1}{20}\frac{(2 n+1) (6 n +1)}{n^2} a_n^{(5)}\ ,\qquad
 b_n^{(1)}=\di\frac{1}{5}\frac{(12 n+5) (4 n +3)}{(4n+1)^2} b_n^{(5)}\ ,
$$
$$
a_n^{(6)}=\di\frac{ a_n^{(5)}}{2(n+1)}\ ,\qquad b_n^{(6)}=\di\frac{2
b_n^{(5)}}{4n+5}\ .
$$

From the recurrent formulas for $a_n^{(5)}$ and $b_n^{(5)}$ it is
possible to derive their expressions via constants $a_0^{(5)} $
and $b_0^{(5)}$:
$$
 a_n^{(5)}=a_0^{(5)}\di\frac{(4 k)^n}
 {n! \cdot \underbrace{3\cdot7\cdot11\cdot\dots(4 n+3)}_
 {\mbox{\small $n$  }}}
$$
$$
 b_n^{(5)}=b_0^{(5)}\di\frac{(4 k)^n}
 {n! \cdot \underbrace{5\cdot9\cdot13\cdot\dots(4 n+5)}_{\mbox{\small $n$ factors}}}
$$

The series determining the solution~(\ref{qe12}) and (\ref{qe13}),
converge at any  $z\ne 0$. Behavior of the moments $M_n $ as
functions of $ \bar z $ are shown in figures Fig 1. Some
discussion of this stationary case that could be considered as the
ground state of Knudsen atmosphere theory is published at
\cite{veresc4}, where a verification of the theoretical results
are made via molecular dynamics simulations. The deviations of
exponential behavior of such atmosphere were discussed also in
\cite{RWV},

{\section{Construction of solutions of the momenta system by
 WKB method.}

 In this section we apply the method WKB
to the  system~(\ref{q7}). We shall assume, that on the bottom
boundary at $z=0$ a wave with characteristic frequency $\omega_0$
is generated. Next we choose the frequency $\omega_0$ to be large
enough, to put characteristic parameter $\xi=\di\frac{3\omega_0
H}{\vT} \gg 1$. We shall search for the solution in the form:
\begin{equation}
\label{qr01}
M_n=\psi_n \exp(i \omega_0 t)+c.c.\ ,
\end{equation}
where, for example, $\psi_1$, corresponding to the moment $M_1$,
is given by the expansion:
\begin{equation}
\label{qr1}
 \psi_1=\sum_{k=1}^6 \sum_{m=1}^\infty \di\frac{1}{(i \xi)^m} A_m^{(k)} \exp(i \xi \varphi_k(z))\ ,
\end{equation}
here $\varphi_k (z)$ - the phase functions corresponding to
different roots of dispersion relation. For other moments $M_n\,,
\quad n=2, \dots, 6$ corresponding functions $\psi_n$ are given by
similar to~(\ref{qr1}) expansion. The appropriate coefficients of
the series we shall designate by corresponding  $B_m^{(k)} \,
C_m^{(k)} \, D_m^{(k)} \, E_m^{(k)} \, F_m^{(k)}$. Substituting
the series~(\ref{qr1}) at the system~(\ref{q7})one arrives at
algebraic equations for the coefficients of  ~(\ref {qr1}) in each
 order. The condition
of solutions existence  results in the mentioned dispersion
relation:
\begin{equation}
\label{qr2}
\begin{array}{l}
 6 {\eta}^3 + (5 u^2 + 20 i u - 21) {\eta}^2 + (5 i u^3 - 24 u^2 - 33 i u + 10) \eta- \vspace{2mm}\\
 - 3 i u^3 + 9 u^2 + 9 i u - 3 = 0\ .
\end{array}
\end{equation}

Here for convenience the following designations are entered:
$$
\left(\frac{\partial \varphi_k}{\partial z}\right)^2=\di\frac{2}{9}\eta_k\ , \qquad
u=\frac{\nu_0}{\omega_0} \exp(-\bar z)\ .
$$
For the coefficients $A_1^{(k)} \, B_1^{(k)} \, \dots $ the
algebraic relations are obtained:
$$
\begin{array}{rcl}
      B_1^{(k)}&=&\mp \di\sqrt{\frac{2}{\eta_k}}A_1^{(k)}\ ,\qquad
      C_1^{(k)}=\di\frac{1}{3}\frac{4 i u-1+\eta_k}{2 i u+3 \eta_k} A_1^{(k)}\ ,\qquad
      D_1^{(k)}=\pm \di\frac{1}{2}\frac{1+5 \eta_k}{2 i u+3 \eta_k} \di\sqrt{\frac{2}{\eta_k}}A_1^{(k)}\ ,\vspace{2mm}\\
      E_1^{(k)}&=&-\di\frac{1}{\eta_k}\frac{3 i u -5 i u \eta_k +5 \eta_k-5 {\eta_k}^2} {2 i u+3 \eta_k} A_1^{(k)}\ ,
      \qquad \quad
      F_1^{(k)}=\mp \di\frac{i \sqrt{2 \eta_k}}{3 u} E_1^{(k)}\ .\\
   \end{array}
$$

The dispersion relation~(\ref {qr2}) represents the cubic equation
with variable coefficients, therefore the exact analytical
solution by formula  Cardano looks very bulky and inconvenient for
 analysis. We study the behavior of solutions at $ \nu \to 0 $ (free
molecular regime) and
 $\nu\to \infty$ (a hydrodynamical regime).

At $\nu=0$ the dispersive relation becomes:
$$
 6 {\eta}^3 -21 {\eta}^2+10 \eta-3=0\ .
$$
The roots are:
$$
 \eta_1 = 3\ ,\quad \eta_2 \approx 0.25 + 0.32 i\ ,\quad
 \eta_3 \approx 0.25 - 0.32 i\ .
$$
Specifying roots~(\ref {qr2}) by the theory of perturbations up to
$u^3$ for the three solutions branches  it is obtained:
$$
\begin{array}{rcl}
\eta_1&=&3 - 0.56 u^2 - {\it i}(1.96 u - 0.17 u^3)\dots\ ,\vspace{2mm}\\
\eta_2&=&(0.25 + 0.32 i) + (0.07 - 0.69 i)u - (0.14 - 0.52 i)u^2 + (0.19 - 0.08 i)u^3\dots\ ,\vspace{2mm} \\
\eta_3&=&(0.25 - 0.32 i) - (0.07 + 0.69 i)u - (0.14 + 0.52 i)u^2 + (0.19 + 0.08 i)u^3\dots\ .\\
\end{array}
$$
Correspondingly for $\varphi_{z k}=\di\frac {\sqrt{2}}{3}
\sqrt{\eta_k}$ we have:
$$
\begin{array}{rcl}
\varphi_{z1}&=&1.15 - 0.47 u^2 - i(0.37 u - 0.17 u^3)\dots\ ,\vspace{2mm}\\
\varphi_{z2}&=&(0.38 + 0.19 i) - (0.12 + 0.34 i)u + (0.12 + 0.02 i)u^2 + (0.06 + 0.07 i)u^3\dots\ ,\vspace{2mm}\\
\varphi_{z3}&=&(0.38 - 0.19 i) + (0.12 - 0.34 i)u + (0.12 - 0.02 i)u^2 - (0.06 - 0.07 i)u^3\dots\ .\\
\end{array}
$$

Similarly in a limit $\nu\to \infty $ (a hydrodynamical limit) for
 solutions of the equation~(\ref{qr2}) $\eta_k$ it is derived:
$$
\begin{array}{l}
\smallskip
\eta_1=0.6 - 1.87 u^{-2} - i(0.72 u^{-1} - 4.27 u^{-3})\ ,\\
\smallskip
\eta_2=- i (u + 0.80 u^{-1}) + 1.40 + 0.18 u^{-2} \ ,\\
\smallskip
\eta_3=- 0.83 u^2 + 1.50 - i(2.33 u - 1.52 u^{-1})\ .\\
\end{array}
$$
The first root relates to the acoustic branch. Accordingly, for
the $\varphi_{zk}$ we have:
$$
\begin{array}{l}
\smallskip
\varphi_{z1}\approx 0.52 - 0.71 u^{-2} - i(0.31 u^{-1} - 1.41 u^{-3})\dots\ ,\\
\smallskip
\varphi_{z2}\approx\sqrt{u} (1-i)(0.47 + 0.30 u^{-2}) +
                   \sqrt{u} (1+i)(0.33 u^{-1} + 0.07 u^{-3})\dots\ ,\\
\smallskip
\varphi_{z3}\approx -0.85 + 0.62 u^{-2} + i(0.61 u  + 0.05 u^{-2})\dots\ .\\
\end{array}
$$

The solution of the equation~(\ref {qr2}) at any $u$ is evaluated
numerically. Behavior of real and imaginary parts $\eta_n$ and
$\varphi_{zk}$ as functions of $u$, and  their behavior on a
complex plane are presented at the figures 1 - 4.

\includegraphics [width=10cm, height=14cm, angle =-90] {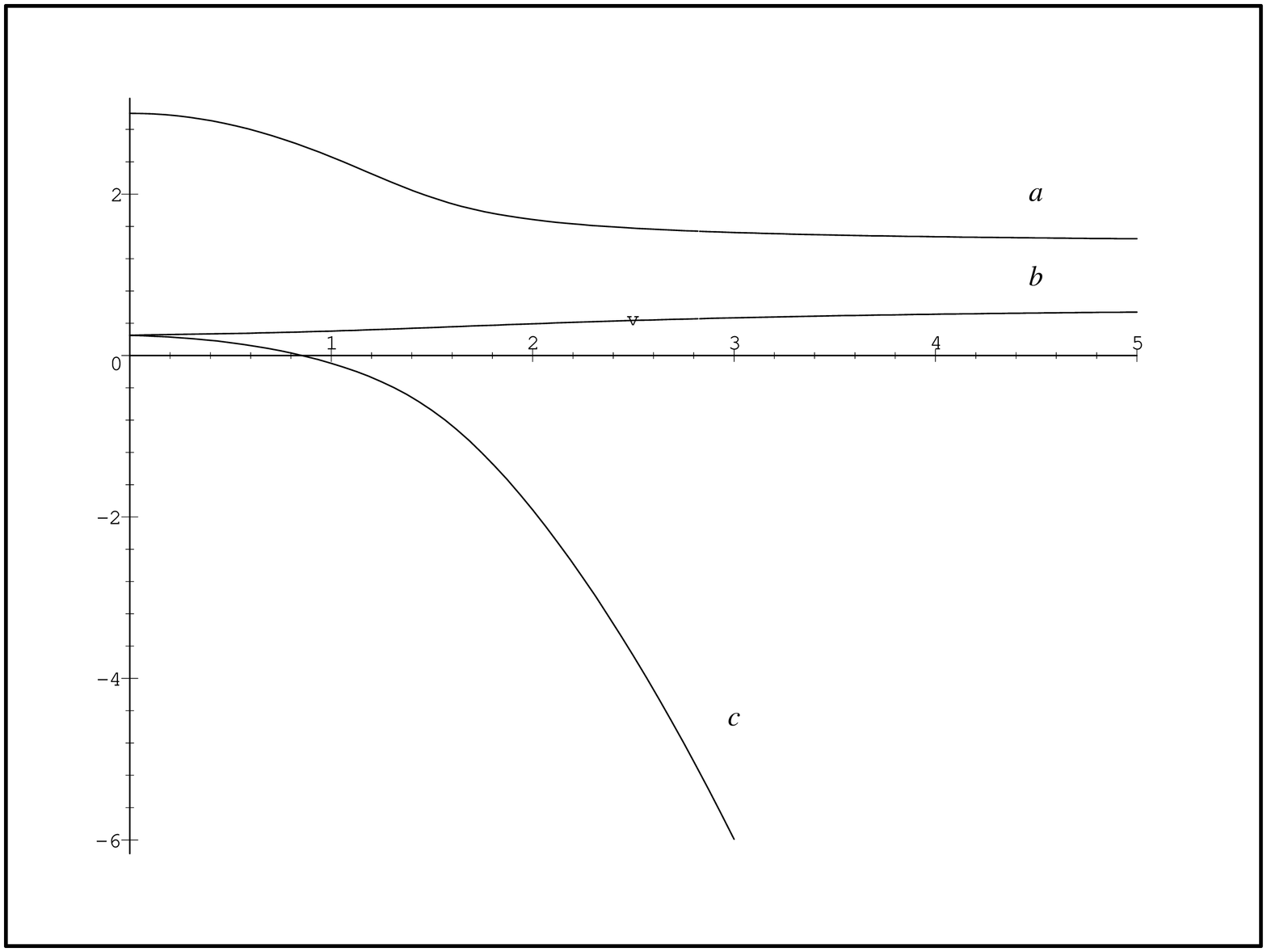}
\begin {center}
Fig. 1. Behavior of the real part of roots of a dispersion
relation:

a - $ \eta_1 $, b - $ \eta_2 $, c - $ \eta_3 $
\end {center}

\includegraphics [width=10cm, height=14cm, angle =-90] {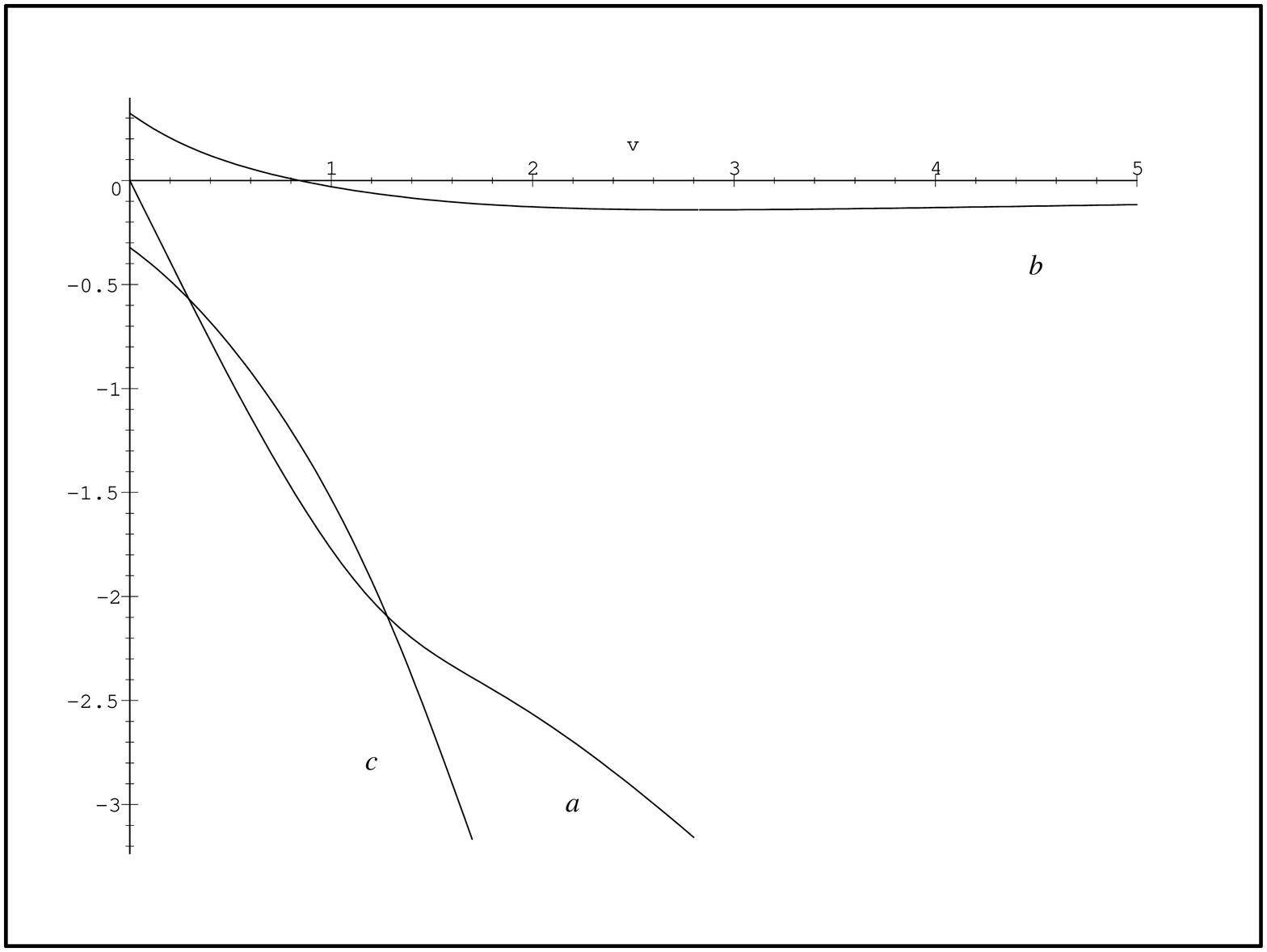}
\begin {center}
Fig. 2. Behavior of the imaginary part of the roots of the
dispersive relation:

a - $ \eta_1 $, b - $ \eta_2 $, c - $ \eta_3 $
\end {center}

\includegraphics [width=10cm, height=14cm, angle =-90] {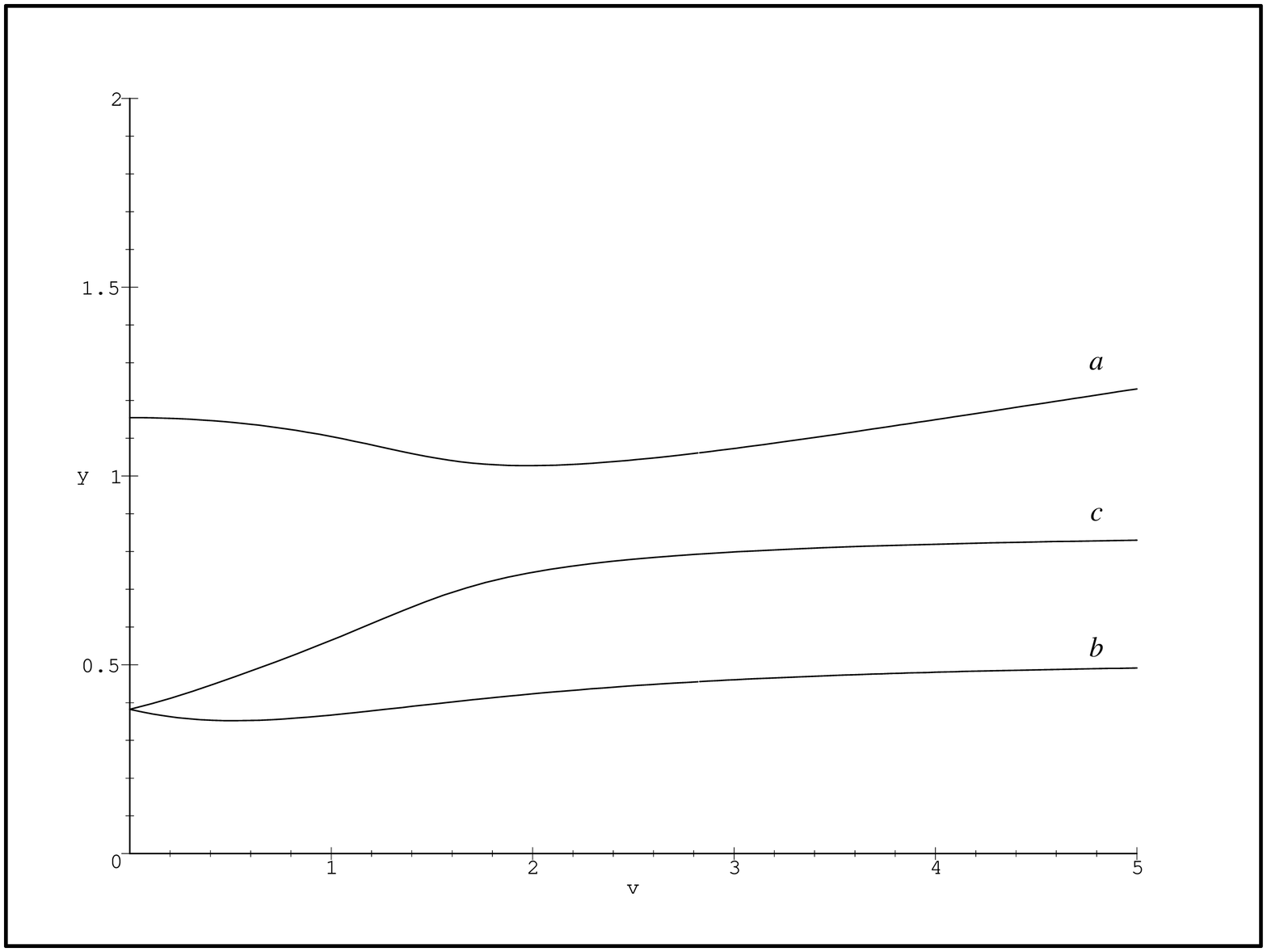}
\begin {center}
Fig. 3. Behavior of the real part of the phase functions $ \phi _
{n z} $: a - $ \phi _ {1z} $, b - $ \phi _ {2z} $, c - $ \phi _
{3z} $
\end {center}

\includegraphics [width=10cm, height=14cm, angle =-90] {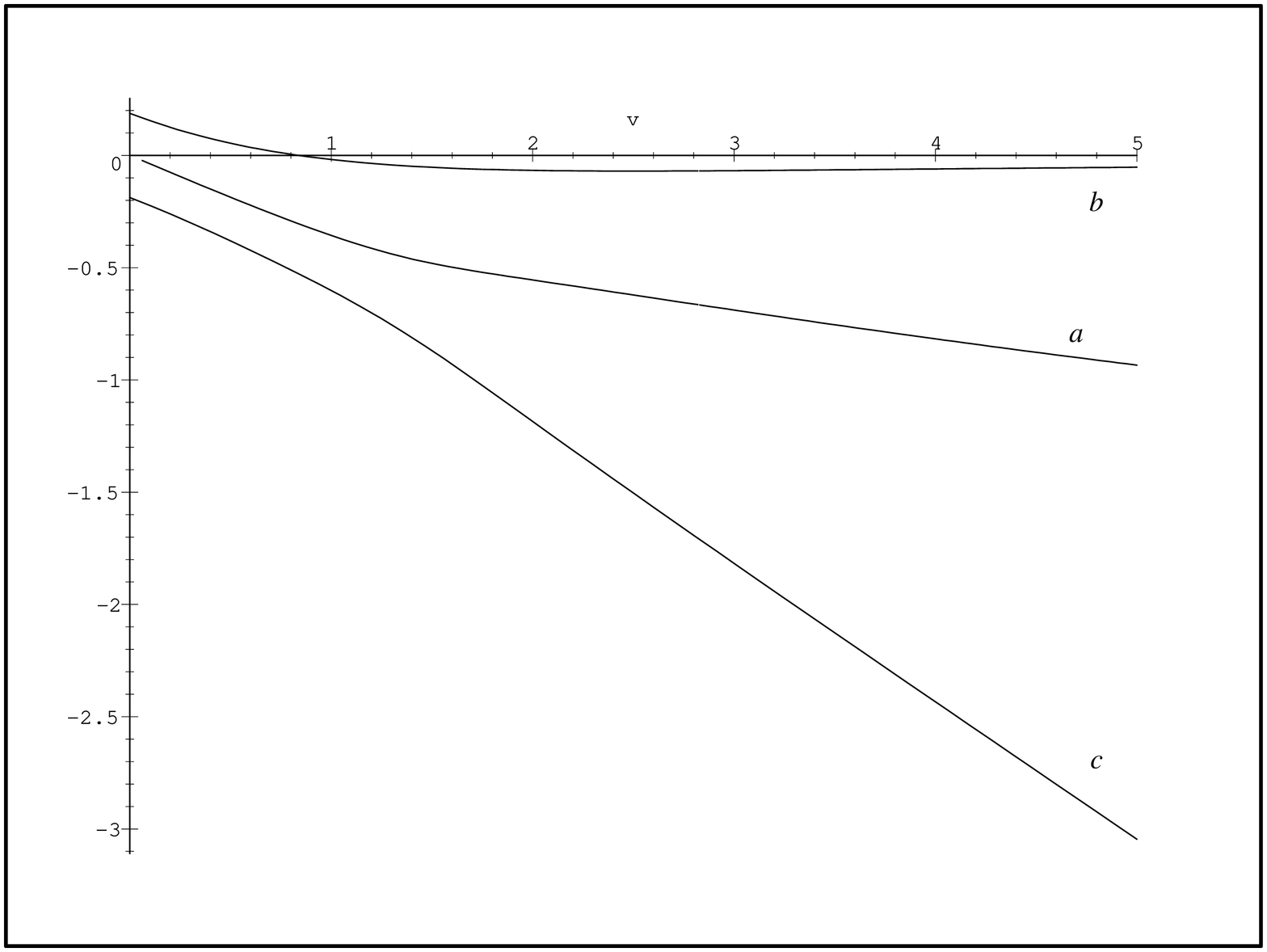}
\begin {center}
Fig. 4. Behavior of the imaginary part of the phase functions $
\phi _ {n z} $:

a - $ \phi _ {1z} $, b - $ \phi _ {2z} $, c - $ \phi _ {3z} $
\end {center}

As an illustration let us consider a problem of generation and
propagation of a gas disturbance, by a plane oscillating with a
given frequency $\omega_0$. Let's assume, that at height $z=0$ all
moments $M_k=0$, except for $M_2=U_0$. In evaluations we
  we   choose the dimensionless frequency equal to
$u(0)=0.1 $, that fits   heights $z~300$ in the Earth atmosphere.
Behavior of the solutions for $M_k$ constructed by the VKB
technique    are shown on fig. 7.

\includegraphics [width=10cm, height=14cm, angle =-90] {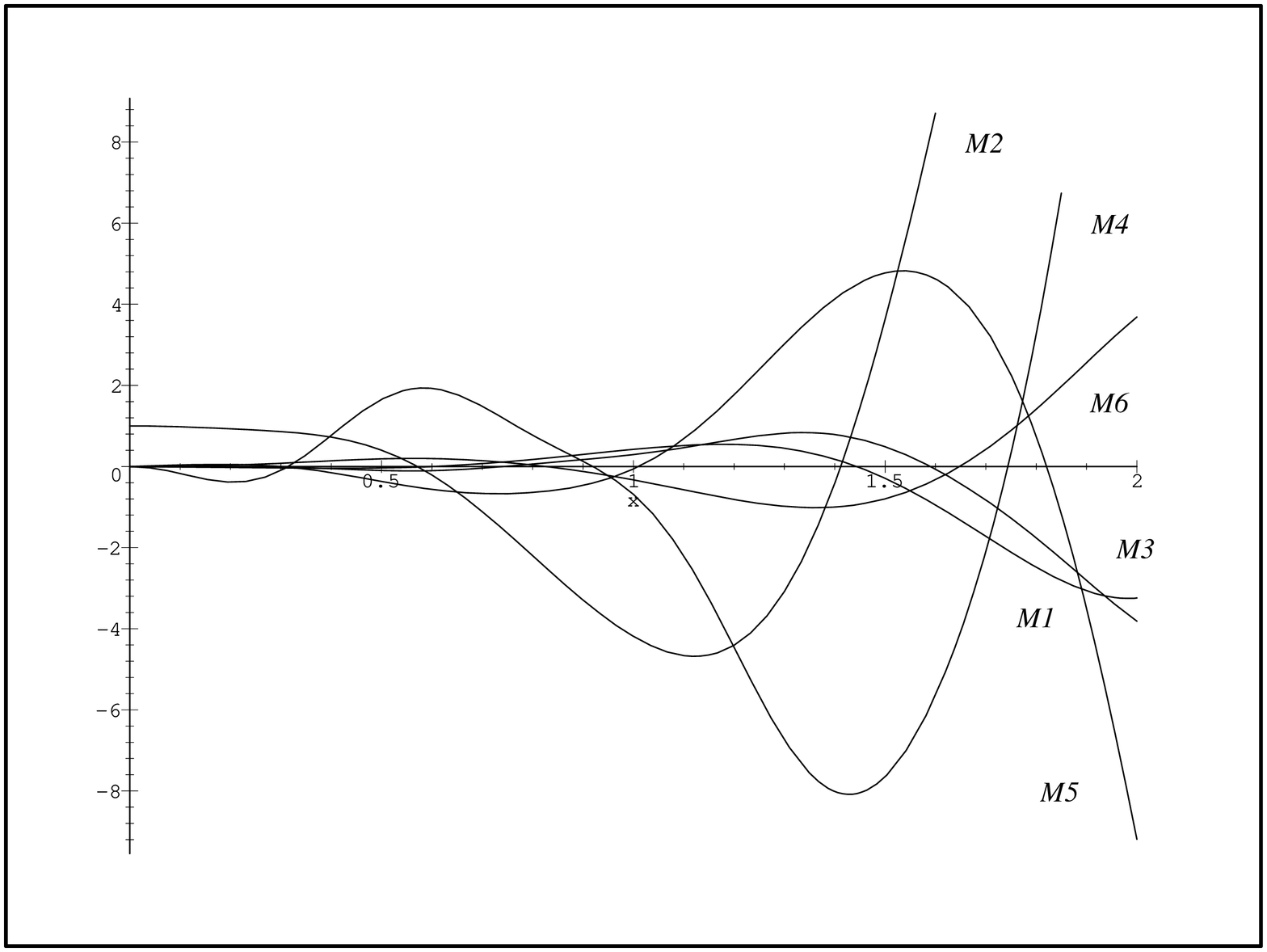}
\begin {center}
Fig. 5. Dependence of the moments $M_k$ on height, $u(0)=0.1 $:
\end {center}
\section{Conclusion}
In this paper we propose a one-dimensional theory of linear
disturbances in a gas, stratified in gravity field, hence
propagating through regions with crucially different Kn numbers.
The regime of the propagation dramatically changes from a
typically hydrodynamic  to the free-molecular one. We also studied
three-dimensional case, to be published elsewhere. The theory is
based on BGK or Gross-Jakson kinetic equation, which solution is
built by means of locally equilibrium distribution function with
different local parameters for molecules moving "up"and "down".
Equations for six moments yields in the closed fluid mechanics system.
For the interesting generalizations of the foundation of such
theory see the recent review of Alexeev \cite{{aleks1}}.
\section{Acknoledgement}
We would like to thank M. Solovchuk for important technical help.

\end {document}